# Integration of short gold nanoparticles chain on SOI waveguide

M. Fevrier* P. Gogol, A. Aassime, R. Mégy, P. Beauvillain, J. M. Lourtioz, and B. Dagens*

Institut d'Electronique Fondamentale, Université Paris Sud, CNRS, UMR 8622, 91405 Orsay, France

**Abstract:** In this letter, we demonstrate the integration of short metal nanoparticle chains (L ≈ 700nm) supporting localized surface plasmons in Silicon On Insulator (SOI) waveguides at telecom wavelengths. Nanoparticles are deposited on the waveguide top and excited through the evanescent field of the TE waveguide modes. Finite difference time domain calculations and waveguide transmission measurements reveal that almost all the TE mode energy can be transferred to nanoparticle chains at resonance. It is also shown that the transmission level is very sensitive to the nanoparticle environment, thus opening the way towards ultra-compact sensors in guided optics on SOI.

* Electronic mail: mickael.fevrier@u-psud.fr, beatrice.dagens@ief.u-psud.fr

Plasmonic waveguides have received much attention in the past few years owing to their ability to spatially confine light well below the diffraction limit [1]. Recent investigations have been carried out to compensate for metal losses using amplifier media [2] – [4] as well as to integrate plasmonic structures in large-scale-integration technology [5] – [11]. Combining plasmonic structures with silicon photonics thus represents an elegant way to bridge the gaps between macroscopic optics and nanodevices with either optical or (opto)electronic functionality. So far, essentially two main types of plasmonic waveguide geometry have been investigated on a silicon-on-insulator (SOI) platform : (i) slot waveguides composed of two metal strips separated by a nanoscale dielectric slot [5] – [9] and (ii) hybrid metal-silicon waveguides consisting of a thin metal film deposited on a narrow SOI waveguide section [10], [11] . Strong confinement of light was indeed demonstrated for both structures by channeling the electromagnetic field of an SOI waveguide mode into a plasmonic guide of reduced dimensions. However, this in turn required strong modifications of the SOI platform as for instance the insertion of either tapered sections or guide interruptions.

In this letter, we demonstrate the integration of another plasmonic waveguide geometry, which consists of a short metal nanoparticle (MNP) chain deposited on top of a SOI waveguide. Energy transfer via dipolar interactions between closely spaced MNPs supporting localized surface plasmons (LSP) leads to the formation of a waveguide that can confine light at smaller scales than previous plasmonic guides [12] – [14]. Because the MNP chain is excited through the evanescent field of the TE waveguide modes, standard SOI waveguides can be used without requiring any specific modification of the SOI structure. Here we show that even in the case of chain lengths as small as ≈ 700 nm (<λ/2), almost all the TE mode energy can be transferred to the MNP chain at resonance. It is also shown that the transmission spectrum of the MNP-loaded SOI waveguide can be significantly modified by slightly changing the external environment of nanoparticles. This opens the way towards ultra-compact bio- or chemical- sensors [15] as well as to optical tweezers [16] in guided optics on SOI. Because of the small size of MNPs, one could envisage the detection of a very small number of molecules of interest.

Fig. 1(a) shows a schematic view of the fabricated structure, which consists of five gold nanoparticles deposited on top of a SOI waveguide. The 500×250nm$^2$ waveguide cross-section allows operating the waveguide on its fundamental TE mode over the spectral range of interest from 1260 to 1630 nm. The size and shape of nanoparticles are accurately determined in such a way that the LSP resonance can be excited by the evanescent field of this mode. The MNP shape is ellipsoidal (fig.1 (b)) with the long axis (D1=203 ± 5nm) parallel to the electric field and the short axis (D2=83 ± 5nm) parallel to the propagation direction of the guided wave. The center-to-center distance between adjacent particles is chosen equal to d=150nm so as to provide a sufficiently strong coupling between metallic dipoles. It is worthwhile noticing that the precision achieved in the fabrication allows us to control the LSP resonance with a wavelength accuracy better than ≈ 15nm. The fabrication of the MNP-loaded SOI waveguide includes two main steps. Firstly, a standard deep-UV lithography process followed by RIE etching and photoresist removal is used to fabricate the SOI waveguide. Secondly, gold nanoparticles are fabricated on top of the waveguide by e-beam lithography and lift-off process. A 30 nm thick gold layer is deposited by e-beam evaporation. A 1 nm titanium (Ti) adhesion layer is deposited prior to the deposition of gold.

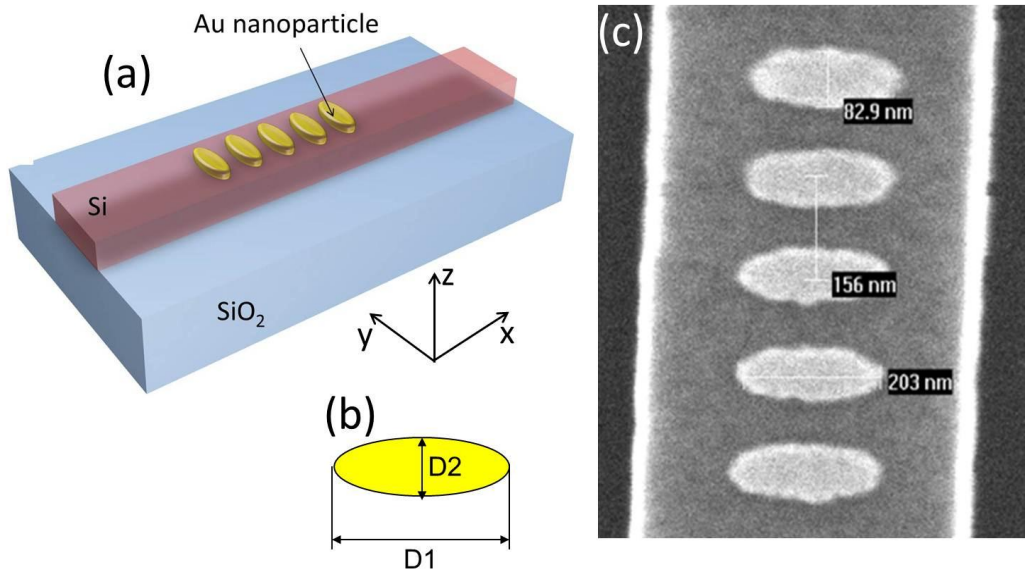

FIG. 1. (a) Schematic view of the structure that we consider. (b) Picture showing the ellipsoidal shape of gold nanoparticles with D1 (respectively D2) the long (respectively small) axis size. (c) Scanning Electron Microscopy image of five gold MNP centered on the of SOI waveguide.

The transmission spectrum of the MNP-loaded SOI waveguide was measured by injecting a wavelength-tunable TE polarized light at the waveguide entrance. A reference waveguide without MNP was used on the same chip for transmission normalization. Both as-cleaved waveguides were ended by tapered sections to optimize the light injection. The input light was delivered by a tunable laser scanned by steps of 1nm over the 1260-1630 nm range. A lensed polarization maintaining fiber was used to couple the laser light to the entrance facet of the SOI waveguide. The light at the sample output was collected by an objective with a × 20 magnification and a 0.35 numerical aperture, and was focused onto a power meter. Fig. 2(a) (blue curve) shows the normalized transmission spectrum measured between 1260 and 1450 nm. A transmission minimum close to zero is obtained at 1325 nm, which corresponds to the maximum excitation of nanoparticles and to the highest ohmic losses. Almost the entire energy of the TE waveguide mode is then transferred to the MNP chain. In contrast, far from LSP resonance, MNPs have a weak influence, and the waveguide transmission approaches 100%. "Noise" on the transmission spectrum is due to Fabry-Perot (FP) oscillations caused by reflections at the Si waveguide facets. For an overall waveguide length of 4 mm, the period of FP oscillations is ≈ 0.06 nm so that they cannot be resolved with a 1nm step scan as used in experiments.

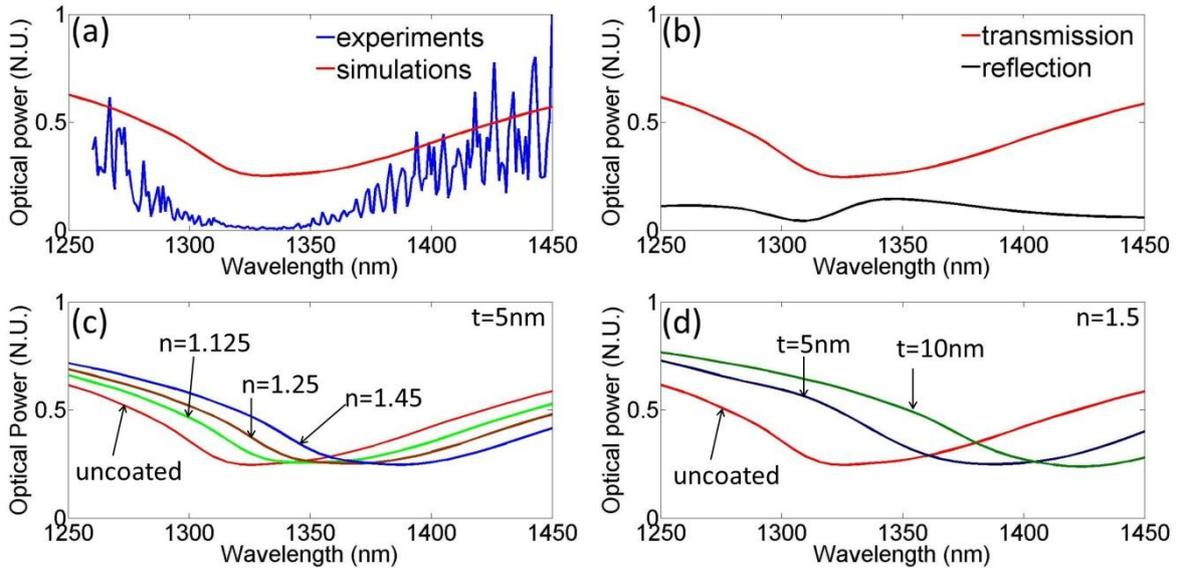

FIG. 2. (a) Normalized transmission spectrum of the SOI waveguide with gold nanoparticles deposited on top. Blue curve is for measurements. Red curve is for FDTD calculations. (b) Calculated reflection (black) and transmission (red) spectra. (c-d) waveguide transmission calculated for gold nanoparticles coated with a lossless dielectric material:(c) 5 nm thick coating with refractive index n=1.125, 1.25 and 1.45 (d) 5 nm and 10 nm thick coatings with n=1.5 refractive index. Red curve in (c) and (d) is the reference for uncoated gold.

A 3D finite-difference time-domain (FDTD) model from Lumerical with a uniform mesh of 3 nm×3 nm×3 nm was used to model the spectral behavior of the fabricated structures. Accurate dispersion data were introduced for deposited gold after fitting a Drude model to experimental ellipsometric measurements. The presence of a thin layer of native oxide between Si and Ti was also accounted for in calculations. In contrast, reflections at the Si waveguide facets were disregarded to avoid extremely time-consuming calculations. Fig. 2(a) (red curve) shows the calculated transmission spectrum, which is in very good agreement with the experimental one. Both the general shape of the spectral response and the position of the transmission minimum are well reproduced by calculations. Fig. 2(b) provides additional information about the reflection spectrum (black curve) calculated from the FDTD model. As seen, the level of reflection never exceeds 15% over the entire spectral range. A dispersion-like evolution is observed around the LSP resonance.

The FDTD model was further exploited to calculate the field intensity along the MNP chain and the SOI waveguide for three wavelengths: 1250, 1325 and 1450 nm. Fig. 3 (left column) shows the field intensity maps along the propagation direction (x-z plane). For each wavelength, a substantial fraction of the SOI waveguide mode intensity is transferred to the MNP chain, which thus forms a short plasmonic waveguide. The propagating wave experiences ohmic losses before re-coupling to the dielectric guide at the end of the chain. In each case, interference patterns are observed in the first part of the guide due to reflections at the guide discontinuity. The three field maps essentially differ from each other by the amount of energy deposited in the chain and by the field distribution along the different particles of the chain. At 1250 nm, the energy transfer from the SOI waveguide to the chain is only partial, thus suggesting that dielectric and plasmonic modes have different wavevectors. The field maximum in the chain is located in the first particles. At 1325 nm (transmission minimum, fig. 2(a)), the energy transfer to the MNP chain is almost total, indicating similar wavevectors for the two modes. The field maximum occurs at the middle of the chain. At 1450 nm, the energy transfer to the chain is decreased, and the field is concentrated in the last particles of the chain. The spatial shift of the field maximum with wavelength presents similarities with that recently reported for a particle array excited by an external plane wave [17]. Fig. 3 (middle column) illustrates previous evolutions from field intensity profiles calculated along the chain axis (blue curve) and along the SOI waveguide axis (red curve). Two field maxima are observed for each particle corresponding to the two air/metal interfaces. Fig. 3 (right column) shows top-view maps of the field intensity calculated in the mid-plane of the MNP chain. Again, previous evolutions are verified for the three wavelengths. More importantly, field maxima are found at the two extremities of each nanoparticle as expected for a dipolar excitation.

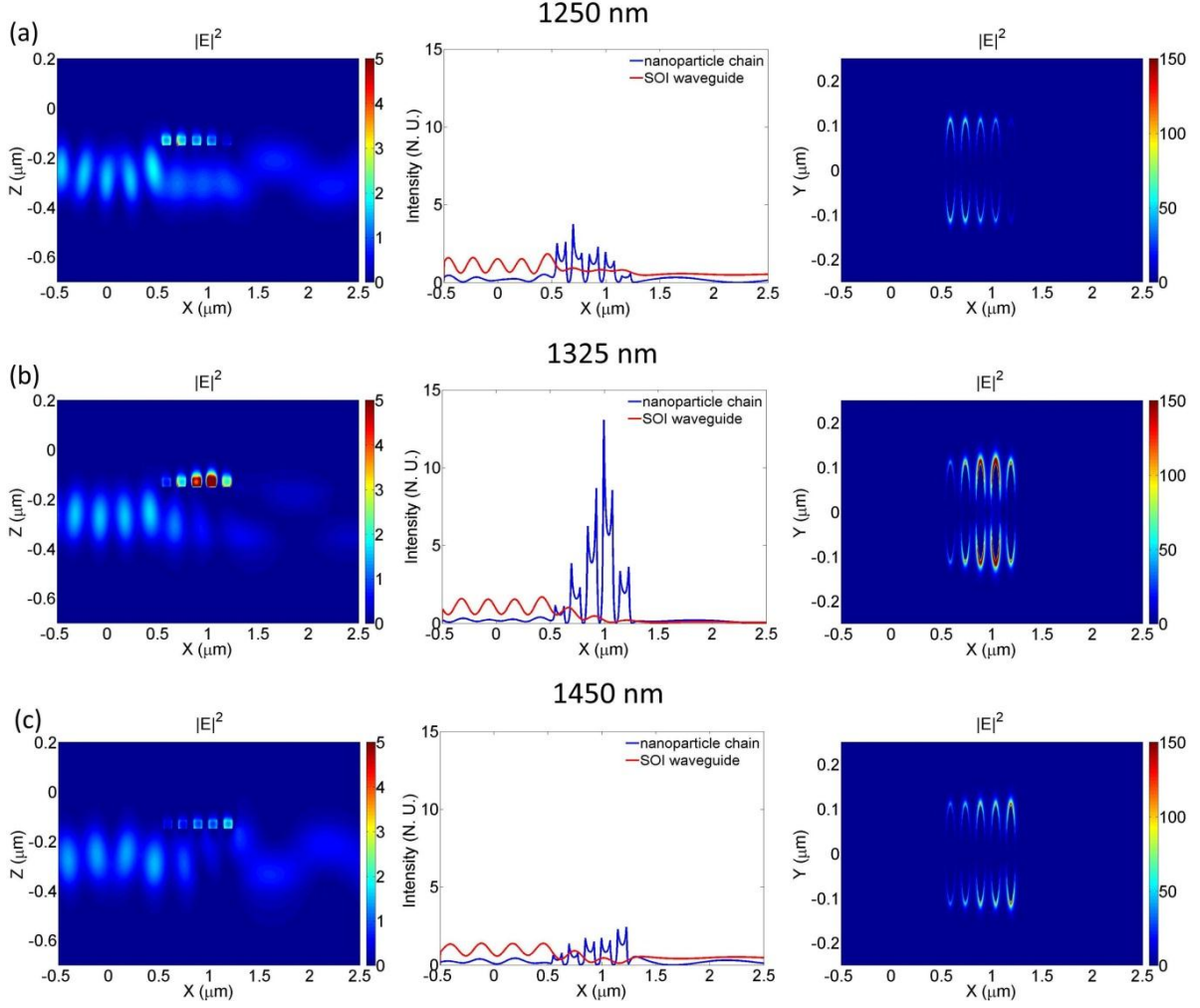

FIG. 3. Left column: maps of the field intensity ($|E|^2$) calculated in the vertical symmetry plane of the structure for three wavelengths, 1250 nm (a), 1325 nm (b) and 1450 nm (c). Middle column: intensity profiles calculated along the nanoparticle chain axis (blue) and the SOI waveguide axis (red). Right column: top view maps of the field intensity ($|E|^2$) calculated in the mid-plane of the nanoparticle chain.

The localization of the field in the MNP chain suggests a strong dependence of the waveguide behavior with the external environment of nanoparticles. The sensing performances of the structure were numerically investigated by artificially incorporating a thin dielectric coating at the air/gold interface of the metallic chain to simulate a molecular environment of particles. As the cubic mesh used in FDTD calculations imposed coating thicknesses of at least 3 nm, smaller thicknesses were simulated by using low values of refractive index. Fig. 2(c-d) represent the evolution of the waveguide transmission for dielectric coatings with a 5 (10) nm thickness and a refractive index varying from 1.125 to 1.45. The latter value corresponds, for instance, to a dielectric medium composed of thiol molecules that are currently used in plasmonic bio-sensors. As seen in fig. 2(c), for a thickness of 5 nm and a refractive index as low as 1.125, the transmission minimum is redshifted by ≈ 25 nm while the transmission value at λ = 1300nm is increased by more than 30%. Such variations should be easily detected in experiments. Previous values of coating parameters approximate a molecular volume of ≈ 0.1 attoliter (≈$10^5$ $nm^3$) for an MNP chain length of ≈ 700 nm as that used in the experiments. For a coating thickness of 10 nm and a refractive index of 1.5 fig.2(d), the wavelength redshift at transmission minimum reaches 100 nm while the transmission value at λ = 1315nm is multiplied by a factor of ≈ 2.5.

In summary, we have reported the fabrication and optical characterization of very short metal nanoparticle chains on standard SOI waveguides at near-infrared wavelengths. Waveguide transmission measurements have been found to be in very good agreement with FDTD calculations, which also revealed the waveguide-like behavior of the chains and their strong coupling to the TE waveguide mode at localized surface plasmon resonance. Almost the entire waveguide mode energy can be transferred to chains as short as 5×d (≈ 700 nm) at this resonance. Thanks to the accuracy achieved in fabrication, the resonance wavelength itself can be controlled with a relative precision better than 5%. These results are believed to be key steps towards the integration of nanometer-size plasmonic functions in silicon photonics. Localized surface plasmons potentially offer a wide variety of guiding configurations since metallic nanoparticles can be arranged on demand on SOI waveguides. Resonant dipole excitation and field confinement in MNPs also open the way to enhanced optical (non-linear) interactions and/or very sensitive sensors. Numerical investigations carried out in this work have shown the high sensitivity of the waveguide structure to the external environment of nanoparticles. The results are believed to be of great promise for the development of ultra-compact sensors integrated in guided optics on SOI.

The authors acknowledge David Bouville for sample preparation. They also thank A. Chelnokov and C. Delacour for fruitful discussion. This work has been supported by the Agence Nationale de la Recherche under contract PLACIDO No ANR-08-BLAN-0285-01. One grant has been funded by Region Ile-de-France.